\documentclass[a4paper]{jpconf}
\usepackage{graphicx}
\usepackage{amssymb}
\begin{document}
\title{From Feynman Proof of Maxwell Equations to Noncommutative Quantum Mechanics}

\author{A. B\'erard$^1$, H. Mohrbach$^1$, J. Lages$^2$, P. Gosselin$^3$, Y. Grandati$^1$, H. Boumrar$^4$, F. M\'enas$^4$}

\address{$^1$ LPMC-ICPMB, FR 2843 du CNRS, Institut de Physique, Universit\'e Paul
Verlaine-Metz, 1 Bd. Arago, 57078 Metz, Cedex 3, France}

\address{$^2$ Institut UTINAM, Dynamique des Structures Complexes, UMR 6213 du CNRS,
Universit\'e de Franche-Comt\'e, 25030 Besan\c{c}on, Cedex, France}

\address{$^3$ Institut Fourier, UMR 5582 du CNRS, UFR de
Math\'ematiques, Universit\'e de Grenoble I, BP74, 38402 St Martin d'H\`eres, Cedex, France}

\address{$^4$ Laboratoire de Physique et de Chimie Quantique, Universit\'e Mouloud Mammeri de Tizi Ouzou, Alg\'erie}

\ead{aberard001@noos.fr}
\ead{jose.lages@utinam.cnrs.fr}

\begin{abstract}
In 1990, Dyson published a proof due to Feynman of the Maxwell
equations assuming only the commutation relations between position
and velocity. With this minimal assumption, Feynman never supposed
the existence of Hamiltonian or Lagrangian formalism. In the
present communication, we review the study of a relativistic
particle using ``Feynman brackets.'' We show that Poincar\'e's
magnetic angular momentum and Dirac magnetic monopole are the
consequences of the structure of the Lorentz Lie algebra defined
by the Feynman's brackets. Then, we extend these ideas to the dual
momentum space by considering noncommutative quantum mechanics. In
this context, we show that the noncommutativity of the coordinates
is responsible for a new effect called the spin Hall effect. We
also show its relation with the Berry phase notion. As a practical
application, we found an unusual spin-orbit contribution of a
nonrelativistic particle that could be experimentally tested.
Another practical application is the Berry phase effect on the
propagation of light in inhomogeneous media.
\end{abstract}

\section{Introduction}

Various ways exist to present the Maxwell equations. The usual one
is the historical approach in which the empirical basis for each
equation is initially given. Another remarkable way is exposed in
an old unpublished work of Feynman, reported in an elegant paper
by Dyson \cite{DYSON} published in 1990. The initial Feynman's
motivation was to develop a quantization procedure without
resorting to a Lagrangian or a Hamiltonian. For this, let s
consider a nonrelativistic particle of mass $m$ subjected to an
external force
$m\displaystyle\frac{d\dot{x}^{i}}{dt}=F^{i}(\mathbf{x},\dot{\mathbf{x}},t)$
and the fiber tangent space with a symplectic structure defined by
the ``Feynman brackets'' $\left[ x^{i},x^{j}\right] =0$ and
$m\left[x^{i},\dot{x}^{j}\right] =\delta ^{ij}$. The Feynman
brackets obey the Leibnitz law
\begin{equation}
\frac{d}{dt}\left[ f(\mathbf{x},\dot{\mathbf{x}}),g(\mathbf{x},%
\dot{\mathbf{x}})\right] =\left[ \frac{df(\mathbf{x},\dot{%
\mathbf{x}})}{dt},g(\mathbf{x},\dot{\mathbf{x}})\right] +\left[ f(%
\mathbf{x},\dot{\mathbf{x}}),\frac{dg(\mathbf{x},\dot{%
\mathbf{x}})}{dt}\right]
\end{equation}
and the Jacobi identity
\begin{equation}
\left[ x^{i},\left[ x^{j},x^{k}\right] \right] +\left[ x^{j},\left[
x^{k},x^{i}\right] \right] +\left[ x^{k},\left[ x^{i},x^{j}\right] \right] =0.
\end{equation}
From these assumptions, Feynman, in $1948$, deduced the following
relations :
\begin{eqnarray}
\left[ \dot{x}^{i},\dot{x}^{j}\right] &=&F^{ij}(\mathbf{x}%
,t)=\varepsilon ^{ijk}B_{k}(\mathbf{x},t) \\
F^{i}(\mathbf{x},\dot{\mathbf{x}},t) &=&E^{i}(\mathbf{x}%
,t)+\varepsilon ^{ijk}\dot{x}_{j}B_{k}(\mathbf{x},t) \\
\mathbf{\nabla}\cdot\mathbf{B} &=&0 \;\;\;\mbox{ and }\;\;\;\mathbf{\nabla}\wedge\mathbf{E}%
=-\frac{\partial\mathbf{B}}{\partial t},
\end{eqnarray}
which actually corresponds to the Lorentz force and the first
group of Maxwell equations. The result seems very strange since
starting with a classical equation (the Newton's law), we end up
with relativistic equations. In fact, ``only'' the first group of
Maxwell equations is retrieved; the second group, according to
Dyson, is a simple definition of matter. This is not a new idea;
Le Bellac and Levy-Leblond \cite{LEBELLAC} have already studied
the Galilean invariance of these equations. Although with this
approach, a Lagrangian or Hamiltonian structure is unnecessary,
Hojman and Shepley \cite{HOJMAN} showed that by using a Helmholtz
inverse variational problem under certain conditions, an action
can be associated to these Feynman commutation relations.

The interpretation of the Feynman's derivation of Maxwell's
equations has generated
\cite{HOJMAN,TANIMURA,NOUS3,LEE,CHOU,CARINENA,HUGHES,MONTESINOS,SINGH,SILAGADZE}
a great interest among physicists. In particular, Tanimura
\cite{TANIMURA} has generalized the Feynman's derivation in a
Lorentz covariant form with a scalar time evolution parameter. An
extension of the Tanimura's approach has been achieved
\cite{NOUS3} using the Hodge duality to derive the two groups of
Maxwell's equations with a magnetic monopole in flat and curved
spaces. In Ref.~\cite{LEE}, the descriptions of relativistic and
nonrelativistic particles in an electromagnetic field were
studied, whereas in Ref.~\cite{CHOU}, a dynamical equation for
spinning particles was proposed. A rigorous mathematical
interpretation of Feynman's derivation associated with the inverse
problem for Poisson dynamics has been formulated in Ref.~\cite
{CARINENA}. Other works \cite{MONTESINOS,SINGH,SILAGADZE} have
provided new insights ino the Feynman's derivation of the
Maxwell's equations. Recently \cite{NOUS6}, some of the authors
embedded Feynman's derivation of the Maxwell's equation in the
framework of noncommutative geometry. As Feynman's brackets can be
interpreted as a deformation of Poisson brackets, we then showed
that the Feynman brackets can be viewed as a generalization of the
Moyal brackets defined over the tangent bundle space. We must also
quote a new and very interesting study in noncommutative space by
Carinena and Figueroa \cite{CARINENA1}.

The noncommutation of the velocities in the presence of an
electromagnetic field implies that the angular algebra symmetry,
\emph{e.g.} the sO(3) symmetry in the Euclidean space, is broken.
If we restore such a symmetry, we point out the necessity of
adding a Poincar\'{e} momentum $\mathbf{M}$ to the simple angular
momentum $\mathbf{L}$. Then, the direct consequence of this
restoration is the generation of a Dirac magnetic monopole. The
extension of these ideas to the covariant case in a Minkowski
space is proposed in Section 2, where we show that this symmetry
induces a magnetic angular momentum \cite{NOUS4} as well as a new
electric angular momentum. In Section 3, we include our work
\cite{NOUS7} in the natural generalization of quantum mechanics
involving noncommutative space-time coordinates. This
generalization was originally introduced by Snyder \cite{SNYDER}
as a short-distance regularization to improve the problem of
infinite self-energies inherent in the quantum field theory. Due
to the advent of the renormalization theory, this idea was not
very popular until Connes \cite{CONNES} analyzed the Yang Mills
theories on noncommutative space. Recently, a correspondence
between a noncommutative gauge theory and a conventional gauge
theory was introduced by Seiberg and Witten \cite {SEIBERG}.
Noncommutative gauge theories were also found as being naturally
related to the string theory and M-theory \cite{KONECHNY}.
Applications of noncommutative theories were also found in
condensed matter physics, for instance, in the quantum Hall effect
\cite{BELLISSARD} and in the noncommutative Landau problem
\cite{JACKIW1,HORVATHY}. Then, the name of noncomutative quantum
mechanics began to appear, notably with the works of
\cite{HORVATHY,GAMBOA}. In the context of noncommutative quantum
mechanics, we present a summary of the paper \cite{NOUS7} where
the restoration of the sO(3) symmetry leads to the introduction of
a dual Dirac monopole in momentum space. This monopole was
recently experimentally found in solid state physics \cite{FANG}.
Finally, from the study of the Dirac equation with an unspecified
potential in an adiabatic approximation, we investigate
\cite{NOUS8} the link that exists between this formalism and the
presence of a Berry phase, which plays an important role in the
definition of the position operator.

\section{Lorentz symmetry with Feynman brackets}

One of the most important symmetry in physics is the spherical
symmetry corresponding to the isotropy of physical space. This
symmetry is related to the sO(3) algebra. We showed within the
Feynamn brackets formalism that this symmetry is broken in the
presence of an electromagnetic field. The restoration of this
symmetry results in a Poincar\'e momentum and a Dirac monopole
\cite{NOUS4}. The generalization to the Lorentz symmetry is direct
by assuming a particle of mass $m$ moving in Minkowski space with
position $x^{\mu }(\tau )$ ($\mu =1,2,3,4)$ depending on parameter
$\tau $ with the following commutation relations :
\begin{equation}
\left[ x^{\mu },\dot{x}^{\nu }\right] =\frac{\eta^{\mu \nu }}{m},
\label{deux}
\end{equation}
where $\eta^{\mu \nu }$ is the metric. Following the same steps as
those described in \cite{NOUS1}, we consider the angular momentum
$L^{\mu\nu}=m(x^\mu\dot{x}^{\nu}-x^{\nu }\dot{x}^{\mu })$. With
the Feynman brackets, we recover the standard Lorentz Lie algebra
and the transformation laws of position and velocity.

We then generalize the Feynman's approach by considering the
following brackets \cite{NOUS2} :
\begin{equation}
\left[ \dot{x}^{\mu },\dot{x}^{\nu }\right] =\frac{1}{m^{2}}%
(qF^{\nu \mu }+g\,^*\!F^{\nu \mu }),  \label{hodge}
\end{equation}
where g is the magnetic charge of the magnetic monopole and the
*-operation is the Hodge duality. The derivative with respect to
the time parameter of Eq. \ref{hodge} leads to the following
equation of motion
\begin{equation}
m\ddot{x}^{\mu }=qF^{\mu \nu
}(\mathbf{x})\dot{x}_{\nu}+g\,^{*}\!F^{\mu \nu
}(\mathbf{x})\dot{x}_{\nu}+G^{\mu }(\mathbf{x}).
\end{equation}
The $G$ field (which has no physical interpretation until now)
also satisfies the equation $\partial ^{\mu }G^{\nu }-\partial
^{\nu }G^{\mu }=0$. In the presence of the gauge fields, the
Lorentz Lie algebra structure becomes more complicated.
\begin{equation}
\left\{
\begin{array}{lll}
\left[ x^{\mu },L^{\rho \sigma }\right]
&=&\eta^{\mu \sigma }x^{\rho }-\eta^{\mu\rho }x^{\sigma } \\
\\
\left[ \dot{x}^{\mu },L^{\rho \sigma }\right]
&=&
\eta^{\mu \sigma }\dot{x}^{\rho }-\eta^{\mu \rho }\dot{x}^{\sigma }+\displaystyle\frac{q}{m}(F^{\mu \sigma }\dot{x}^{\rho }-F^{\mu \rho }\dot{x}^{\sigma})
+\displaystyle\frac{g}{m}(^{*}\!F^{\mu \sigma }\dot{x}^{\rho }-{}^{*}\!F^{\mu \rho }\dot{x}^{\sigma }) \\
\\
\left[ L^{\mu \nu },L^{\rho \sigma }\right]
&=&
\eta^{\mu \rho }L^{\nu \sigma}-\eta^{\nu \rho }L^{\mu \sigma }+\eta^{\mu \sigma }L^{\rho \nu }-\eta^{\nu \sigma}L^{\rho \mu }\\
&&
+q(x^{\mu }x^{\rho }F^{\nu \sigma }-x^{\nu }x^{\rho }F^{\mu \sigma }+x^{\mu
}x^{\sigma }F^{\rho \nu }-x^{\nu }x^{\sigma }F^{\rho \mu }) \\
&&
+g(^{*}\!F^{\nu \sigma }x^{\mu }x^{\rho }-{}^{*}\!F^{\mu \sigma }x^{\nu }x^{\rho
}+{}^{*}\!F^{\rho \nu }x^{\mu }x^{\sigma }-{}^{*}\!F^{\rho \mu }x^{\nu }x^{\sigma })
\end{array}
\right.
\end{equation}
Therefore, we introduce a generalized electromagnetic angular
momentum $\mathcal{L}^{\mu \nu }=L^{\mu \nu }+M^{\mu \nu }$ in
order to recover the usual algebra in the absence of
electromagnetic fields. The requirement that the generalized
electromagnetic angular momentum satisfies the usual algebra
imposes constraints on the tensor $M$ that can be solved leading
to the following results
\begin{equation}
M^{ij}=q(F^{ij}x^{k}x_{k}-F^{j}{}_{k}x^{k}x^{i}-F_{k}{}^{i}x^{k}x^{j})+g({}^*\!F^{ij}x^{k}x_{k}-{}^*\!F^{j}{}_{k}x^{k}x^{i}-{}^*\!F_{k}{}^{i}x^{k}x^{j})
\end{equation}
for the space components. The new angular momentum is, therefore,
the sum of two contributions, a magnetic and an electric one :
\begin{equation}
\mathbf{M}=-q(\mathbf{x}\cdot\mathbf{B})\mathbf{x}+g(\mathbf{x}\cdot\mathbf{E})\mathbf{x}=\mathbf{M}_{m}+\mathbf{M}_{e}=-(\mathbf{x}\cdot\mathbf{P})\mathbf{x},
\label{vingt deux}
\end{equation}
where $\mathbf{M}_{m}=-q(\mathbf{x}\cdot\mathbf{B})\mathbf{x}$ and
$\mathbf{M}_{e}=g(\mathbf{x}\cdot\mathbf{E})\mathbf{x}$ are the
magnetic and electric angular momenta and
$\mathbf{P}=q\mathbf{B}-g\mathbf{E}$.

Now, we require the Jacobi identity between the velocities,
$\left[ \dot{x}^{\mu },\left[ \dot{x}^{\nu },\dot{x}^{\rho
}\right]
\right] +\left[ \dot{x}^{\nu },\left[ \dot{x}^{\rho },%
\dot{x}^{\mu }\right] \right] +\left[ \dot{x}^{\rho },\left[
\dot{x}^{\mu },\dot{x}^{\nu }\right] \right] =0$ which yields the
generalized Maxwell equations \cite{NOUS2}
\begin{equation}
q(\partial ^{\mu }F^{\nu \rho }+\partial ^{\nu }F^{\rho \mu
}+\partial ^{\rho }F^{\mu \nu })+g(\partial^{\mu }\,^*\!F^{\nu
\rho }+\partial ^{\nu }\,^{*}\!F^{\rho \mu }+\partial ^{\rho
}\,^*\!F^{\mu \nu })=0.  \label{quatorze}
\end{equation}
The projection of Eq. \ref{quatorze} on a three-dimensional space gives $%
q\mathbf{\nabla}\cdot\mathbf{B}-g\mathbf{\nabla}\cdot\mathbf{E}=\mathbf{\nabla}\cdot\mathbf{P}=0$,
where $\mathbf{P}$ can be considered either perpendicular to the
vector $\mathbf{x}$ or null. In both these cases, we have
$\mathbf{M}=0$. The Jacobi identity implies either there are no
electric and magnetic monopoles or the two monopoles exactly
compensate each other.

To break this duality symmetry, we no more require the Jacobi
identity and introduce the tensor $N^{\mu\nu\rho}$ as
\begin{equation}
q(\partial ^{\mu }F^{\nu \rho }+\partial ^{\nu }F^{\rho \mu
}+\partial ^{\rho }F^{\mu \nu })+g(\partial ^{\mu }\,^*\!F^{\nu
\rho }+\partial ^{\nu }\,^*\!F^{\rho \mu }+\partial ^{\rho
}\,^*\!F^{\mu \nu })=qgN^{\mu \nu \rho },  \label{dixsept}
\end{equation}
which implies that $\mathbf{\nabla}\cdot\mathbf{P}\neq 0 $. The
constraints on the Lie algebra can be fulfilled by a radial vector
field centered at the origin solution:
$\displaystyle\mathbf{P}=\frac{\mathbf{x}}{4\pi
\|\mathbf{x}\|^{3}}$. As a consequence, we get a nonvanishing
electromagnetic angular momentum
\begin{equation}
\mathbf{M}=\mathbf{M}_{m}+\mathbf{M}_{e}=-\frac{qg}{4\pi }\frac{\mathbf{x}}{\|\mathbf{x}\|}.
\end{equation}
As the modulus of this momentum is radially constant, the magnetic
and electric charges are not independent. It is the famous Dirac
quantization condition connecting these two charges. It clearly
appears that the vector $\mathbf{P}=q\mathbf{B}-g\mathbf{E}$ plays
the same role as the magnetic field in the case without a dual
field or in the three-dimensional theory.

Due to the fact that for these monopoles, the source of the fields
is localized at the origin, we have
\begin{eqnarray}
\mathbf{\nabla}\cdot\mathbf{P} &=&[\dot{x}^{i},[\dot{x}^{j},\dot{x}^{k}]]+[\dot{x}^{j},[\dot{x}^{k},\dot{x}^{i}]]+[\dot{x}^{k},[\dot{x}^{i},\dot{x}^{j}]] \\
&=&q\mathbf{\nabla}\cdot\mathbf{B}-g\mathbf{\nabla}\cdot\mathbf{E}=\frac{qg}{4\pi }\left[x^{l},\frac{x_{l}}{\|\mathbf{x}\|^3}\right]
=qg\delta^{3}(\mathbf{x}).
\end{eqnarray}
For example, we can select $\mathbf{B}=\displaystyle\frac{g}{8\pi
}\displaystyle\frac{\mathbf{x}}{\|\mathbf{x}\|}$ and
$\mathbf{E}=-\displaystyle\frac{q}{8\pi
}\displaystyle\frac{\mathbf{x}}{\|\mathbf{x}\|}$. We have found
that $\mathbf{M}$ is the new angular momentum, which is the sum of
the Poincar\'{e} magnetic angular momentum \cite{POINCARE} plus an
electric angular momentum; $\mathbf{B}$ being the field of a Dirac
magnetic monopole \cite{DIRAC} and $\mathbf{E}$ the electric field
of an electric Coulomb monopole.

In addition, we remark that the generalized angular moment
$\mathcal{L}=m(\mathbf{x}\wedge\mathbf{\dot{x}})-(\mathbf{x}\cdot\mathbf{P})\cdot\mathbf{x}$
is conserved because the particle satisfies the usual equation of
motion.

In conclusion of this section, it is interesting to note that with
this same formalism, we can retrieve two groups of Maxwell
equations and that this procedure of symmetry restoration has also
been performed for Lorentz algebra in a curved space \cite{NOUS6}.
Another generalization of this approach can be found in a recent
interesting work where the study of the Lorentz generators in
N-dimensional Minkowski space has been proposed \cite{LAND}.

\section{Noncommutative quantum mechanics}

Let the momentum vector $\mathbf{p}$ replace the velocity vector $\dot{%
\mathbf{x}}$ in the Feynman formalism presented before. Consider a quantum
particle of mass $m$ whose coordinates satisfy the deformed Heisenberg
algebra
\[
\left[ x^{i},x^{j}\right] =i\hbar q_{\theta }\theta ^{ij}(\mathbf{x},\mathbf{%
\ p}),\;\;\;\left[ x^{i},p^{j}\right] =i\hbar \delta ^{ij},\;\;\;\left[
p^{i},p^{j}\right] =0,
\]
where $\theta $ is a field that is \emph{a priori} position- and
momentum-dependent and $q_{\theta }$ is a charge characterizing
the intensity of the interaction of the particle and the $\theta $
field. The commutation of the momentum implies that there is no
external magnetic field. It is well known that these commutation
relations can be obtained from the deformation of the Poisson
algebra of classical observable with a provided Weyl-Wigner-Moyal
product \cite{MOYAL} expanded at the first order in $\theta $.

\subsection{Dual Dirac monopole in momentum space}

The Jacobi identity $J(p^{i},x^{j},x^{k})=0$ implies an important
property that the $\theta $ field is position-independent
$\theta^{jk}(\mathbf{p})$. Then, one can consider the $\theta$
field as the dual of a magnetic field and $q_{\theta }$ as the
dual of an electric charge. The fact that the field is homogeneous
in space is an essential property for vacuum. In addition, one
easily see that a particle in this field moves freely,
\textit{i.e.}, the vacuum field does not act on the motion of the
particle in the absence of an external potential. The effect of
the $\theta $ field is manifested only in the presence of a
position-dependent potential. To look further at the properties of
the $\theta $ field, consider the other Jacobi identity between
the positions. We then have the equation of motion of the field
\begin{equation}\label{jxxx}
\frac{\partial \theta ^{jk}(\mathbf{p})}{\partial p^{i}}+\frac{\partial
\theta ^{ki}(\mathbf{p})}{\partial p^{j}}+\frac{\partial \theta ^{ij}(%
\mathbf{p})}{\partial p^{k}}=0,  \label{div}
\end{equation}
which is the dual equation of the Maxwell equation $\mathbf{\nabla }\cdot
\mathbf{B}=0$. As we will see later, equation (\ref{div}) is not satisfied
in the presence of a monopole and this will have important consequences.

If we consider the position transformation
\begin{equation}
X^{i}=x^{i}+q_{\theta }a_{\theta }^{i}(\mathbf{x},\mathbf{p}),
\label{position}
\end{equation}
where $a_{\theta }$ is \textit{a priori} position and momentum
dependent, we are able to restore the usual canonical Heisenberg
algebra
\begin{equation}
\left[ X^{i},X^{j}\right] =0,\;\;\;\left[ X^{i},p^{j}\right] =i\hbar \delta
^{ij},\;\;\;\left[ p^{i},p^{j}\right] =0.
\end{equation}
The second commutation relation implies that $a_{\theta }$ is
position-independent. The first commutation relation leads to the
following expression of $\theta $ in terms of the dual gauge field $%
a_{\theta }$ :
\begin{equation}
\theta ^{ij}(\mathbf{p})=\frac{\partial a_{\theta }^{i}(\mathbf{p})}{%
\partial p^{j}}-\frac{\partial a_{\theta }^{j}(\mathbf{p})}{\partial p^{i}},
\label{theta}
\end{equation}
which is  dual of the standard electromagnetic relation in
position space.

Consider now the problem of angular momentum. It is obvious that
the angular momentum expressed according to the canonical
coordinates satisfies the angular momentum algebra; however, it is
not conserved,
\begin{equation}
\frac{d\mathbf{\mathcal{L}}(\mathbf{X},\mathbf{p})}{dt}=kq_{\theta }%
\mathbf{\mathcal{L}}\wedge \mathbf{\Theta }.
\end{equation}
In the original $\left( \mathbf{x},\mathbf{p}\right)$ space, the
usual angular momentum $\ L^{i}(\mathbf{x},\mathbf{p})=\varepsilon
^{i}{}_{jk}x^{j}p^{k}$ does not satisfy this algebra. Therefore,
it seems that there are no rotation generators in the $\left(
\mathbf{x},\mathbf{p}\right) $ space. We will now prove that a
genuine angular momentum can be defined only if $\theta $ is a
nonconstant field.

Indeed, from the definition of angular momentum, we deduce the
following commutation relations :
\begin{equation}
\left\{
\begin{array}{lll}
\lbrack x^{i},L^{j}]&=&i\hbar \varepsilon ^{ijk}x_{k}+i\hbar q_{\theta
}\varepsilon ^{j}{}_{kl}p^{l}\theta ^{ik}(\mathbf{p}),\\
\lbrack p^{i},L^{j}]&=&i\hbar \varepsilon ^{ijk}p_{k},\\
\lbrack L^{i},L^{j}]&=&i\hbar \varepsilon ^{ij}{}_{k}L^{k}+i\hbar q_{\theta
}\varepsilon ^{i}{}_{kl}\varepsilon ^{j}{}_{mn}p^{l}p^{n}\theta ^{km}(\mathbf{p}),
\end{array}
\right.
\end{equation}
particularly showing that the sO(3) algebra is broken. To restore
the angular momentum algebra, consider the transformation law
\begin{equation}
L^{i}\rightarrow \Bbb{L}^{i}=L^{i}+M_{\theta }^{i}(\mathbf{x,p});
\end{equation}
then, the usual sO(3) algebra needs to be satisfied. Then, it can
be shown that this condition can be fulfilled by a dual Dirac
monopole defined in momentum space\footnote{This result has been
already found in an other context \cite{Ska,Blachan}.}
\begin{equation}
\mathbf{\Theta }(\mathbf{p})=\frac{g_{\theta }}{4\pi }\frac{%
\mathbf{p}}{\|\mathbf{p}\|^{3}},  \label{monopole}
\end{equation}
where we introduced the dual magnetic charge $g_{\theta }$
associated with the $\Theta $ field.
Consequently, we have $\mathbf{M}_{\theta }(\mathbf{p%
})=-\displaystyle\frac{q_{\theta }g_{\theta }}{4\pi
}\frac{\mathbf{p}}{\|\mathbf{p}\|} $ , which is the dual of the
famous Poincar\'{e} momentum introduced in position space. Then,
the generalized angular momentum
\begin{equation}
\mathbf{\Bbb{L}}=\mathbf{x}\wedge \mathbf{p} -\frac{q_{\theta
}g_{\theta }}{4\pi }\frac{\mathbf{p}}{\|\mathbf{p}\|}
\end{equation}
is a genuine angular momentum satisfying the usual algebra. One can check
that it is a conserved quantity for a free particle.

The duality between the monopole in momentum space and the Dirac
monopole is due to the symmetry of the commutation relations in
noncommutative quantum mechanics, where $\left[ x^{i},x^{j}\right]
=i\hbar q_{\theta }\varepsilon ^{ijk}\Theta _{k}(\mathbf{p})$, and
the usual quantum mechanics in a magnetic field, where
$\left[\dot{x}^{i},\dot{x}^{j}\right] =i\hbar q\varepsilon
^{ijk}B_{k}( \mathbf{x})$. Therefore, the two gauge fields $\Theta
(\mathbf{p})$ and $B( \mathbf{x})$ are a dual of each other.

Note that in the presence of the dual monopole, the Jacobi
identity (\ref{jxxx}) does not hold
\begin{equation}
\left[ x^{i},\left[ x^{j},x^{k}\right] \right] +\left[ x^{j},\left[
x^{k},x^{i}\right] \right] +\left[ x^{k},\left[ x^{i},x^{j}\right] \right]
=-q_{\theta }\hbar ^{2}\frac{\partial \Theta ^{i}(\mathbf{p})}{\partial p_{i}%
}=-4\pi q_{\theta }\hbar ^{2}g_{\theta }\delta ^{3}(\mathbf{p}).
\end{equation}
This term is responsible for the violation of the associativity,
which is only restored if the following quantification equation is
satisfied : $\int d^{3}p\displaystyle\frac{\partial \Theta
^{i}}{\partial p_{i}}=\displaystyle\frac{2\pi n\hbar }{q_{\theta
}}$, leading to $q_{\theta }g_{\theta }=\displaystyle\frac{n\hbar
}{2}$, which is in complete analogy with Dirac's quantization. We
should also note that a new insight into fuzzy monopoles in the
context of a nonconstant noncommutativity in 2D field theories has
very recently been developed \cite{STERN}.

\subsection{Link with the Berry phase}

In quantum mechanics, this construction may look formal because it
is always possible to introduce commuting coordinates with the
help of the transformation $\mathbf{R=x-p\wedge
S/}p^{2}\mathbf{.}$ The angular momentum is then
$\mathbf{J=R\wedge p+S}$ and it satisfies the usual algebra,
whereas the potential energy term in the Hamiltonian becomes $V$
$(\mathbf{R+p\wedge S/}p^{2})$, which contains spin-orbit
interactions. In fact, the inverse procedure is usually more
efficient. If we consider a Hamiltonian with a particular
spin-orbit interaction, a trivial Hamiltonian with a nontrivial
dynamics due to the noncommutative coordinates algebra can be
obtained. This procedure has been applied with success to the
study of adiabatic transport in semiconductors with spin-orbit
couplings \cite{FANG}. The difficulty is now to decide which one
of the two position operators $\mathbf{x}$ or $\mathbf{R}$ gives
rise to the real mean trajectory of the particle. In fact, it is
well known that $\mathbf{R}$ does not have the correct properties
of a position operator for a relativistic particle. As we shall
see, this point is very important when considering the
nonrelativistic limit as we predict an effect similar to the
Thomas precession but with regard to the velocity.

It should also be noted that other recent theoretical works
concerning the anomalous Hall effect in two-dimensional
ferromagnets predict a topological singularity in the Brillouin
zone \cite{ONODA}. In addition, a monopole in the crystal momentum
space was experimentally discovered and interpreted in terms of an
Abelian Berry curvature \cite{FANG}.

\subsubsection{Dirac equation}

The Dirac's Hamiltonian for a relativistic particle of mass $m$
has the form $\hat{H}=\mathbf{\alpha .p}+\beta m+\hat{V}\left(
\mathbf{R}\right) ,$ where $\hat{V}$ is an operator that acts only
on the orbital degrees of freedom. Using the Foldy-Wouthuysen
unitary transformation, we get the following transformed
Hamiltonian
\begin{equation}
U(\mathbf{p})\hat{H}U(\mathbf{p})^{+}=E_{p}\beta +U(\mathbf{p})\hat{V}%
(i\hbar \partial _{\mathbf{p}})U(\mathbf{p})^{+}.
\end{equation}
The kinetic energy is now diagonal, whereas the potential term becomes $\hat{V%
}(\mathbf{D})$ with the covariant derivative defined by $\mathbf{\ D=}i\hbar
\partial _{\mathbf{p}}+\mathbf{A}$ and with the gauge potential $\mathbf{A}%
=i\hbar U(\mathbf{p})\partial _{\mathbf{p}}U(\mathbf{p})^{+}$. We
now consider the adiabatic approximation that neglects the
interband transition. We then keep only the block diagonal matrix
element in the gauge potential and project it on the subspace of
positive energy. This projection cancels the zitterbewegung, which
corresponds to an oscillatory movement around the mean position of
the particle that combines the positive and negative energies. In
this way, we obtain a nontrivial gauge connection allowing us to
define a new position operator $\mathbf{r}$ for this particle
\begin{equation}
\mathbf{r=}i\hslash \partial _{\mathbf{p}}+i\hbar \mathcal{P}(U\partial _{%
\mathbf{p}}U^{+}),  \label{r}
\end{equation}
where $\mathcal{P}$ is a projector on the positive energy
subspace. It is straightforward to prove that the anomalous part
of the position operator can be interpreted as a Berry connection
in momentum space. In this context, the $\theta $ field we
postulated in \cite{NOUS7} appears as a consequence of the
adiabatic motion of a Dirac particle and corresponds to a
non-Abelian gauge curvature satisfying the relation
\begin{equation}
\theta ^{ij}(\mathbf{p,\sigma })=\partial _{p^{i}}A^{j}-\partial
_{p^{j}}A^{i}+\left[ A^{i},A^{j}\right] .  \label{theta NA}
\end{equation}
The commutation relations between the coordinates are then
expressed as
\begin{equation}
\left[ x^{i},x^{j}\right] =i\hslash \theta
^{ij}(\mathbf{p},\mathbf{\sigma }), \label{nc}
\end{equation}
which has very important consequences as it implies the
nonlocalizability of the spinning particles. This is an intrinsic
property and is not related to the creation of a pair during the
measurement process (for a detailed discussion of this very
important point, see \cite{BACRY}).

Note that it is possible to generalize the construction of the
position operator for a particle with unspecified $n/2$ $(n>1)$
spin through the Bargmann-Wigner equations. In this way, the
general position operator $\mathbf{r}$ for spinning particles is
\begin{equation}
\mathbf{r=}i\hslash \partial _{\mathbf{p}}+\frac{c^{2}\left( \mathbf{p}%
\wedge \mathbf{S}\right) }{E_{p}\left( E_{p}+mc^{2}\right) }.
\label{rs}
\end{equation}
The generalization of (\ref{nc}) is then
\begin{equation}
\left[ x^{i},x^{j}\right] =i\hslash \theta ^{ij}(\mathbf{p},\mathbf{S}%
)=-i\hbar \varepsilon _{ijk}\frac{c^{4}}{E_{p}^{3}}\left( mS^{k}+\frac{p^{k}(%
\mathbf{p}.\mathbf{S)}}{E_{p}+mc^{2}}\right) .
\end{equation}
For a massless particle, we recover the relation
$\mathbf{r=}i\hslash
\partial _{\mathbf{p}}+\mathbf{p}\wedge \mathbf{S/}p^{2}$ with the
commutation relation giving rise to the monopole $\left[ x^{i},x^{j}\right]
=i\hslash \theta ^{ij}(\mathbf{p})=-i\hbar \varepsilon _{ijk}\lambda \frac{%
p^{k}}{p^{3}}$. The momentum space monopole we introduced in
\cite{NOUS7} in order to construct a genuine angular momenta has a
very simple physical interpretation. It corresponds to the Berry
curvature resulting from an adiabatic process of massless particle
with helicity $\lambda $. It is not surprising that a massless
particle has a monopole Berry curvature as it is well known that
the band-touching point acts as a monopole in momentum space
\cite{BERRY}. This is precisely the case for massless particles
for which the positive and negative energy bands are degenerate in
$p=0$. The monopole appears as a limiting case of a more general
non-Abelian Berry curvature arising from an adiabatic process of
massive spinning particles. For $\lambda =\pm 1$, we have the
position operator of the photon, whose noncommutativity property
agrees with the weak localizability of the photon, which is
certainly an experimental fact.

The spin-orbit coupling term in (\ref{rs}) is a very small
correction to the usual operator in the particle physics context,
but it may be strongly enhanced and observed in solid state
physics because the spin-orbit effect is much more important than
in the vacuum. For instance, in narrow-gap semiconductors, the
equations of the bands theory are similar to the Dirac equation
with the forbidden gap $E_{G}$ between the valence and conduction
bands instead of the Dirac gap $2mc^{2}$ \cite{RASHBA2}. The
monopole in momentum space predicted and observed in
semiconductors results from the limit of the vanishing gap
$E_{G}\rightarrow 0$ between the valence and conduction bands.

It is also interesting to look at the symmetry properties of the
position operator with respect to the group of spatial rotations.
In terms of commutative coordinates $\mathbf{R}$, the angular
momentum is by definition $\mathbf{J}=\mathbf{R\wedge
p}+\mathbf{S}$, whereas in terms of the noncommutative
coordinates, the angular momentum reads $\mathbf{J}=\mathbf{\
r\wedge p}+\mathbf{M,}$ where
\begin{equation}
\mathbf{M}=\mathbf{S-A\wedge p.}  \label{m}
\end{equation}
One can explicitly check that in terms of the noncommutative
coordinates, the relation $[x^{i},J^{j}]=i\hbar \varepsilon
^{ijk}x_{k}$ is satisfied; therefore $\mathbf{r}$, like
$\mathbf{R}$, transforms as a vector under space rotations, but
$d\mathbf{R}/dt=c\mathbf{\alpha }$ is physically unacceptable. For
a massless particle, Eq. \ref{m} leads to the Poincar\'{e}
momentum associated with the monopole in momentum space deduced in
\cite{NOUS7}.

\subsubsection{Physical applications}

We are interested to look at some physical properties of the
noncommuting position operator. Let us consider the equation of
motion of a particle in an arbitrary potential. Due to the Berry
phase in the definition of position, the equation of motion will
change. In order to compute a commutator like
$\left[x^{k},V(x)\right]$, one can consider the semiclassical
approximation $\left[ x^{k},V(x)\right] =i\hbar \partial
_{l}V(x)\theta ^{kl}+O(\hbar ^{2})$, which gives the following
equations of motions
\begin{equation}
\dot{\mathbf{r}}=\frac{\mathbf{p}}{E_{p}}-\dot{\mathbf{p}}%
\mathbf{\wedge \theta }\mbox{, \qquad and \qquad }\dot{\mathbf{p}}=-%
\mathbf{\nabla }V\mathbf{(r)}  \label{rp}
\end{equation}
with $\theta ^{i}=\varepsilon ^{ijk}\theta _{jk}/2$. These
equations are the relativistic generalization of the equations
found in \cite{FANG} that leads to the spin Hall effect in the
context of semiconductors. Equations \ref{rp} have also the same
form as those found in \cite{NIU} in the context of condensed
matter physics for a particle (without spin) propagating in a
periodic potential. The difference in this case relies on the fact
that the two Berry phases have a completely different physical
origin as that in \cite{NIU}; the Berry phase is only due to the
periodic potential.

An important physical application of our theory concerns the nonrelativistic
limit of a charged spinning Dirac particle in a potential $\hat{V}(\mathbf{r}%
)$. In this limit, we obtain
\begin{equation}
\widetilde{H}(\mathbf{R,p})\approx mc^{2}+\frac{p^{2}}{2m}+\hat{V}(\mathbf{R}%
)+\frac{e\hbar }{4m^{2}c^{2}}\mathbf{\sigma}\cdot\left( \mathbf{\nabla }\hat{V}(%
\mathbf{r})\mathbf{\wedge p}\right),  \label{H}
\end{equation}
which is a Pauli Hamiltonian with a spin-orbit interaction term.
As shown in \cite{MATHUR}, the Born-Oppenheimer approximation of
the Dirac equation leads to the same nonrelativistic Hamiltonian
as a consequence of the nonrelativistic Berry phase $\theta
^{ij}=-\varepsilon _{ijk}\sigma ^{k}/2mc^{2}$ . Note that in
\cite{MATHUR}, it was also proved that the
adiabaticity condition (which neglects the non-diagonal matrix elements of $%
\hat{V}$ ) is satisfied for slowly varying potentials as long as
$L\gg\lambda $, where $L$ is the length scale over which
$\hat{V}(\mathbf{r})$ varies and $\lambda $ is the de Broglie
wavelength of the particle. As a consequence of Eq. \ref{H}, we
deduce the velocity associated with the usual
Galilean-Schr\"{o}dinger position operator $\mathbf{R}$
\begin{equation}
\frac{dX^{i}}{dt}=\frac{p^{i}}{m}+\frac{e\hslash }{4m^{2}c^{2}}\varepsilon
^{ijk}\sigma _{j}\partial _{k}\hat{V}(\mathbf{r}),  \label{xnr1}
\end{equation}
whereas the nonrelativistic limit of Eq. \ref{rp} leads to the
following velocity operator
\begin{equation}
\frac{dx^{i}}{dt}=\frac{p^{i}}{m}+\frac{e\hslash
}{2m^{2}c^{2}}\varepsilon ^{ijk}\sigma _{j}\partial
_{k}\hat{V}(\mathbf{r});  \label{xnr2}
\end{equation}
this result predicts an enhancement of the spin-orbit coupling
when the new position operator is considered. One can appreciate
the similarity between our result and the Thomas precession.
Indeed, this result offers another manifestation beside the Thomas
precession of the difference between the Galilean limit (leading
to Eq. \ref{xnr1}) and the nonrelativistic limit (leading to Eq.
\ref{xnr2}).

The ultrarelativistic limit gives us another example of
topological spin transport. Experimentally, a topological spin
transport has already been observed in the case of the photon
propagation in an inhomogeneous medium \cite{ZELDOVICH}, where the
right and left circular polarization propagate along different
trajectories in a waveguide (the transverse shift is observable
due to the multiple reflections), a phenomenon interpreted quantum
mechanically as arising from the interaction between the orbital
momentum and the spin of the photon \cite{ZELDOVICH}. To interpret
the experiments, these authors introduced a complicated
phenomenological Hamiltonian. Our approach provides a new
satisfactory interpretation as this effect, also called the
optical Magnus effect, is now explained in terms of the
noncommutative property of the position operator that contains the
spin-orbit interaction. In this sense, this effect is just the
ultrarelativistic spin Hall effect. Note that the adiabaticity
criteria has been proved to be valid in \cite{CHAO}. To illustrate
our purpose, consider the simple photon Hamiltonian in the
inhomogeneous medium $H=pc/n(r).$ The equations of motion
$\dot{x}=\displaystyle\frac{1}{i\hbar }\left[ x,H\right] $ and
$\dot{p}=\displaystyle\frac{1}{i\hbar }\left[ p,H\right] $ in the
semiclassical approximation leads to following relation between
velocity and momentum
\begin{equation}
\frac{dx^{i}}{dt}=\frac{c}{n}\left( \frac{p^{i}}{p}+\frac{\lambda
\varepsilon ^{ijk}p_{k}}{p^{2}}\frac{\partial \ln n}{\partial
x^{j}}\right), \label{nopt}
\end{equation}
which contains an unusual contribution due to the Berry phase. As
a consequence, the velocity is no more equal to $c/n$. Equations
\ref{nopt} are the same as those introduced phenomenologically in
\cite{ZELDOVICH}, but here, they are rigorously deduced from
different physical consideration. Similar equations are also given
in \cite{BLIOKH} where the optical Magnus effect is also
interpreted in terms of a monopole Berry curvature, but in the
context of geometrical optics. Our theory is generalizable to the
photon propagation in a nonisotropic medium; a situation that is
mentioned in \cite{ZELDOVICH}, but could not be studied with their
phenomenological approach.

\section{Conclusion}

In this communication, we have tried to stress the close link,
between the Feynman's approach of the Maxwell equations and
noncommutative quantum mechanics. In particular, we show that the
restoration of broken symmetries leads to the appearance of Dirac
monopoles either in configuration space or momentum space.
Actually, the generalization of Feynman's ideas to the case of the
noncommutative quantum mechanics is interesting because it
naturally leads to the promotion of the $\theta $ parameter to a
$\theta (\mathbf{p})$ field. As a physical realization of the
noncommutative theory, we showed that the $\theta (\mathbf{p})$
field can be interpreted in some circumstances as a Berry
curvature associated with a Berry phase expressed in momentum
space. This was shown in the context of Dirac particle and photon
propagation. Particulary important is the fact that the Berry
phase leads to a new physically interesting effect called spin
Hall effect.

Recently, noncommutative quantum mechanics has been the topic of
several other works in particle and condensed matter physics. For
instance, it was shown that the Berry phase (which is a spin-orbit
interaction) of the photon influences the geometric optics
equations leading to the Magnus effect of light \cite{NOUS8}.
Actually, the spin-orbit contribution on the propagation of light
has led to a generalization of geometric optics called geometric
spinoptics \cite{HORVATHY1}. Other applications concern the
propagation of Dirac particles and photons in a static
gravitational field \cite{NOUS11,NOUS12} and the semiclassical
equations of motion of electrons in a solid
\cite{NOUS9,NOUS10,NIU}. In semiconductor physics, it has also
been found that a noncommutative geometry underlies the
semiclassical dynamics of electrons in an external electric field.
Here, the noncommutativity property is again the consequence of a
Berry phase inducing a purely topological and dissipationless spin
current (intrinsic spin Hall effect) \cite{FANG}. More generally,
we mention the current efforts carried out in order to better
understand the close relation existing between the noncommutative
geometry and the geometric phase \cite{GHOSH}.

\end{document}